*Spin waves and spin instabilities in quantum plasmas*
Andreev P. A., Kuz'menkov L. S.
Lomonosov Moscow State University, Russia

## Abstract

We describe main ideas of method of many-particle quantum hydrodynamics allows to derive equations for description of quantum plasma evolution. We also present definitions of collective quantum variables suitable for quantum plasmas. We show that evolution of magnetic moments (spins) in quantum plasmas leads to several new branches of wave dispersion: spin-electromagnetic plasma waves and self-consistent spin waves. Propagation of neutron beams through quantum plasmas is also considered. Instabilities appearing due to interaction of magnetic moments of neutrons with plasma are described.

In a gas of neutral classic particles reveals one type of collective excitations having linear dependence of frequency on module of wave vector. Plasmas show to be more complicate system revealing three types of collective excitations even in absence of external fields. This complexity related to a long-range interaction between charged particles. Inserting plasma in an external homogeneous magnetic field we find it become anisotropic medium and number of branches of dispersion dependence increases significantly. All of it is about classic plasmas, but then we get quantum plasmas including dynamic of particle spins, we find additional branches of waves [1]-[3]. Mathematically grounds for appearing of new wave solutions arise due to presence of additional equation in the set of hydrodynamic equations: magnetic moment (spin) evolution equation.

Furthermore, new wave solutions are not all we can get studying evolution of spin in the quantum plasmas. Effect of interaction of neutron beams with quantum plasmas was theoretically discovered. This effect exists due to the spin-spin and spin-current interactions between spins of the beam and spins and electrical currents of plasmas.

For derivation of equations for collective evolution of quantum plasmas we should consider system of many charged quantum particles governed by the many particles Schrodinger equation. Using many-particle wave function of 3N coordinates of N particles and time we should construct a suitable collective variable. Knowledge of classic hydrodynamics gives us a hint. We should take concentration of particles as the first collective variable. Defining operator of the particle concentration as the classic microscopic concentration, where coordinates of particles replaced by operator of coordinates

$$\hat{n} = \sum_{p=1}^{N} \delta(\mathbf{r} - \mathbf{r}_p). \quad (1)$$

Applying quantum mechanical averaging to get microscopic quantum concentration we find [4]-[6]

$$n(\mathbf{r},t) = \int \Psi^+(R,t) \sum_{p=1}^{N} \delta(\mathbf{r} - \hat{\mathbf{r}}_p) \Psi(R,t) dR, \quad (2)$$

where $R = \{\mathbf{r}_1, \mathbf{r}_2, ..., \mathbf{r}_N\}$ is the assembly of coordinate of particles and $dR = \prod_{p=1}^{N} d\mathbf{r}_p$ is the element of volume in 3N dimensional configurational space.

At derivation of QHD equations we need to use the Schrodinger equation $i\hbar \frac{\partial \Psi}{\partial t} = \hat{H} \Psi(R,t)$ and explicit form of the Hamiltonian

$$\hat{H} = \sum_p \left( \frac{1}{2m_p} \hat{\mathbf{D}}_p^2 + e_p \phi_p^{ext} - \gamma_p \hat{\boldsymbol{\sigma}}_p \mathbf{B}_{p(ext)} \right) + \frac{1}{2} \sum_{p,n, p \neq n} \left( e_p e_n G_{pn} - \gamma_p \gamma_n G_{pn}^{\alpha\beta} \hat{\sigma}_p^\alpha \hat{\sigma}_n^\beta \right), \quad (3)$$

where $\hat{\mathbf{D}}_p = -i\hbar\nabla_p - \frac{e_p}{c}\mathbf{A}_{p(ext)}$, particle $\phi_p^{ext}$, $\mathbf{A}_{p(ext)}$ are the scalar and vector potentials of an external electromagnetic field, giving electric field $\mathbf{E}_{p(ext)} = -\nabla_p \phi_p^{ext} - \mathbf{A}_{p(ext)}/c$ and magnetic field $\mathbf{B}_{p(ext)} = \nabla_p \times \mathbf{A}_{p(ext)}$, and $\hat{\boldsymbol{\sigma}}_p$ are the Pauli matrixes. First group of terms describes particles interacting with the external electromagnetic field, including energy of magnetic moments in external magnetic field presented by the third term in the first group. The second group of terms consists of two terms describing interparticle interaction. We included two types of interaction: the Coulomb interaction between charges, and the spin-spin interaction between quasi-static magnetic moments. Therefore we have following quantities: $G_{pn} = \frac{1}{|\mathbf{r}_{pn}|}$ is the Green function of Coulomb interaction,

$$G_{pn}^{\mu\nu} = 4\pi\delta^{\mu\nu}\delta(\mathbf{r}_{pn}) + \nabla_p^\mu \nabla_p^\nu \frac{1}{|\mathbf{r}_{pn}|}$$
$$= -\frac{\delta^{\mu\nu}}{|\mathbf{r}_{pn}|^3} + 3\frac{r_{pn}^\mu r_{pn}^\nu}{|\mathbf{r}_{pn}|^5} + \frac{8\pi}{3}\delta^{\mu\nu}\delta(\mathbf{r}_{pn}) \quad (4)$$

is the Green function of spin-spin interaction. We have presented it in two forms. On the left-hand side we present rather compact form of the Green function. On the right-hand side we present more familiar form of dipole-dipole interaction. To get explicit form we took second space derivative of $1/|\mathbf{r}_{pn}|$. We accounted that it included term proportional to the delta function. Therefore coefficients in front of the delta functions are different at different representations of the Green function.

Considering time evolution of the quantum concentration of particles (2) using the Schrodinger equation with Hamiltonian (3) we get set of general quantum hydrodynamic equations. Truncating set of equations keeping the continuity, Euler, magnetic moment evolution equations and using the self-consistent field approximation we obtain closed set of quantum hydrodynamic equations [3]-[6]. This set is similar by form to set of quantum hydrodynamic equations appearing from the Pauli equation for one particle in an external field. This set can be used for studying of different phenomenon in quantum plasmas of spin-½ particles [3].

Using the set of quantum hydrodynamic equations we studied evolution of small perturbations in system of two species, electrons and ions, in an external magnetic field. Plasma waves are deeply related to dynamic of electric field in the plasma. Therefore we expressed all perturbations in terms of the electric field perturbations and obtained a dispersion equation. Solving this equation we analytically got quantum amendments to some classic dispersion dependences. New terms in dispersion equation related to spin of particles lead to appearance of new wave solution. We found two solutions propagating parallel to external field, one for electrons, another one, having similar form, for ions. Four solutions (two for electrons, two for ions) we found considering wave propagations perpendicular to external magnetic field. Existence of these waves is caused by spin of particles. They appear as evolution of electromagnetic perturbations. So we can call them spin-electromagnetic plasma waves.

We also considered specific type of waves. Let us focus our attention on spin (magnetic moment) evolution. If we have system of chargeless magnetic moments we see that they interact by magnetic field created by magnetic moments. This interaction gives possibility for spin-wave propagation. So we considered possibility of similar in system of charged spinning particles. To extract these particular waves we consider waves propagating by means of perturbations of magnetic field (self-consistent spin waves). We assume that the electric field existing in plasma gives no contribution in mechanism of propagation of these spin waves. Thus we put the electric field equals to zero in the set of quantum hydrodynamic equations

$\mathbf{E} = 0$. Considering propagation of the spin waves parallel to the external magnetic field we find three solutions. Two of them have constant frequencies $\omega \approx |\Omega_e| = |eB_0/mc|$, and $\omega \approx |\Omega_i| = |eB_0/mc|$. The third solution has hybrid nature. It exists in systems of two species (at least two species) [3]. Its' dispersion dependence has form

$$\omega^2 = k^2 \frac{v_{Qi}^2 \cdot \chi_e |\Omega_e| - v_{Qe}^2 \cdot \chi_i \Omega_i}{\chi_e |\Omega_e| - \chi_i \Omega_i}. \quad (5)$$

where $\chi_a = M_{0a}/B_{ext}$, $M_{0a}$ is the equilibrium magnetization, $v_{Qa}^2 = v_{Fa}^2/3 + \hbar^2 k^2/(4m_a^2)$. From formula (5) we can see that this solution does not exist in electron-positron systems, since for positrons we have $v_{Qe}^2 = v_{Qp}^2$, $\chi_e = \chi_p$, $|\Omega_e| = |\Omega_p|$.

Oblique propagation of the spin waves was also considered (see Ref. [3] for some details).

Effect of instabilities appearing at an electron beam propagation through plasmas. This effect appears due to the Coulomb interaction between charges of the beam and charges of the plasma. Similar effect can be found due to the spin-spin and spin-current interactions between a beam of neutral magnetised particles (neutrons for instance) *and* spins and electrical currents in plasma.

Different instabilities at resonances of a beam branch with different branches of dispersion dependences of plasma waves were considered in Ref. [3].

Let us to illustrate it on resonance of a beam mode with the electron electromagnetic-spin wave propagating perpendicular to the external magnetic field having frequency approximately equals to the electron cyclotron frequency. The instability exists when monoenergetic beam propagate perpendicular to the external field either [3]. Frequency shift of the wave due to presence of the beam appears as

$$\frac{\delta \omega}{|\Omega_e|} \simeq \pm 2\pi i \sqrt{\chi_b \chi_e \frac{|\Omega_b|}{|\Omega_e|}} \frac{1}{1 + \frac{m_i}{m_e} \frac{v_A^2}{c^2}}, \quad (6)$$

where $v_A = B_0/\sqrt{4\pi n_0 m_i}$ is the Alfven velocity.

As a conclusion we point out major results of this talk. We have considered linear properties of spin-1/2 quantum plasmas. We have demonstrated that spin evolution leads to amendments to classic wave dispersion, new branches of wave dispersion, "spin" waves propagating by means of magnetic field, instabilities caused by neutron beam via the spin-spin and spin-current interactions.